\documentclass{revtex4}

\usepackage{graphicx}
\usepackage{subeqnarray}
\usepackage{hyperref}

\begin{document}

\title{\Large On the Letelier spacetime with quintessence: solution, thermodynamics and Hawking radiation\footnote{Preprint of an article submitted for consideration in the International Journal of Modern Physics D © 2018 [copyright World Scientific Publishing Company] [https://www.worldscientific.com/worldscinet/ijmpd]}}
\date{\today}

\author{Marco M. Dias e Costa}
\email{marco@ifpb.edu.br}
\author{J. M. Toledo}
\email{jefferson.m.toledo@gmail.com}
\author{ V. B. Bezerra}
\email{valdir@fisica.ufpb.br}  
\affiliation{Departamento de Física, Universidade Federal da Para\'iba,
João Pessoa, PB, Brasil, Caixa Postal 5008}
\thispagestyle{empty}

\begin{abstract}
We obtain the solution corresponding to a static and spherically symmetric black hole with a cloud of strings (Letelier spacetime) immersed in a quintessential fluid. We discuss some aspects of its thermodynamics. We also present a discussion about Hawking radiation of particles, in the background under consideration.
\end{abstract}



\maketitle

\section{Introduction}

    In the later 1970's, Letelier \cite{letelier1979clouds} introduced a gauge invariant model of a cloud of strings with the aim to treat gravity coupled to these array of strings, in the framework of general relativity. Thus, using the cloud formed by strings as a source of the gravitational field, Letelier obtained a class of solutions of the Einstein equations corresponding to plane-symmetric, spherically and cylindrically symmetric spacetimes \cite{letelier1979clouds}. In the case of spherical symmetry, he obtained a generalization of the Schwarzschild solution to the one corresponding to a black hole surrounded by a spherically symmetric cloud of strings, which is known as Letelier spacetime \cite{barbosa2016rotating}.  Later on, Letelier used the same model to do an extension \cite{letelier1981fluids}, which includes the pressure, and thus a fluid of strings is considered rather than a cloud. In this case, the general solution for a fluid of strings with spherical symmetry was obtained.

The main motivation to construct those models was based on the fact that the universe can be represented, in principle, by a collection of extended objects, like one-dimensional strings, rather than of point particles, in a more appropriate way.

    Recent theoretical developments suggest that it is necessary to consider extended objects because they offered a potential alternative to be used as the fundamental elements to describe physical phenomena which occur in the universe. From the gravitational point of view, it is important to investigate, for example, a black hole immersed in a cloud or fluid of strings \cite{letelier1979clouds}, due to the fact that these sources have astrophysical observable consequences \cite{kar1997stringy,larsen1997string,soleng1995dark}.

    Studies concerning different aspects associated with the physics of a cloud of strings\cite{soleng1995dark,stachel1980thickening}, and a fluid of strings \cite{smalley1997spinning}, in the framework of general relativity, as well as, in modified theories of gravity \cite{mazharimousavi2010theorem,herscovich2010black,ghosh2014cloud,lee2015lovelock} have been performed during the last decades.

Recent theoretical and experimental researches related to the structure and evolution of the universe indicate that, besides the baryonic and dark matters, there exists a kind of cosmic energy responsible for the accelerated expansion of the observed Universe, whose origin and nature is still unknown. Such an exotic energy, also called dark energy, should be the main constituent of the universe in order to account for its flatness. There are models based on a cosmological constant, where dark energy has a constant density throughout the universe, as well as there are those ones in which this density varies, as in a scenario with the presence of quintessence\cite{caldwell2000introduction}. 

It behaves like the inflaton, which is often modeled as a scalar field slowly rolling down a potential barrier. At cosmological scales, quintessence can cause important effects which can be used to explain the accelerated expansion of our Universe \cite{caldwell2000introduction}. At astrophysical scales, quintessence can cause remarkable effects as, for example, the gravitational deflection of the light coming from distant stars \cite{liu2009influence}. These effects can be more appropriately studied in the context of the general relativity by means of some solutions, like the one corresponding to a black hole surrounded by quintessence, obtained by Kiselev \cite{kiselev2003quintessence}, recently.

Thus, considering extended objects corresponding to one-dimensional strings and the quintessence as sources of the gravitational field, we investigate the solution corresponding to a static and spherically symmetric black hole with a cloud of strings (Letelier spacetime) immersed in a quintessential fluid. Then, we discuss the horizons and the thermodynamics following some results in literature \cite{thomas2012thermodynamics,ghaderi2016thermodynamics} related to the study of black holes surrounded by quintessence. We also present a discussion about Hawking radiation \cite{hawking1975particle} of particles, in the background under consideration. In both studies, we emphasize the role played by the cloud of strings as well as by the quintessence.

This paper is organized as follows. In section \ref{sec2}, we present brief reviews concerning the solutions corresponding to a static and spherically symmetric black hole with a cloud of strings (Letelier spacetime) and a static and spherically symmetric black hole immersed in a quintessential fluid (Kiselev spacetime). In section \ref{sec3}, we present the solution corresponding to a black hole with a cloud of strings and surrounded by quintessence (Letelier spacetime with quintessence). In sections \ref{sec4} and \ref{sec5}, we investigate the horizons and the thermodynamics of this black hole. In section \ref{sec6}, we study the Hawking radiation. Finally, section \ref{sec7} is devoted to the concluding remarks.

\section{Letelier and Kiselev spacetimes}
\label{sec2}

In this section, we shall briefly review the solution of the Einstein equations corresponding to a black hole with cloud of strings with spherical symmetry (Letelier spacetime) \cite{barbosa2016rotating} and the solution which corresponds to a black hole surrounded by quintessence (Kiselev spacetime) \cite{kiselev2003quintessence}.

\subsection{Letelier spacetime}

A moving infinitesimally thin string traces out a two-dimensional world sheet, $\Sigma$, which can be described by the equation

\begin{equation}
x^\mu = x^\mu (\xi^a), \qquad a=0,1,
\end{equation} 

\noindent where $\xi^0$ and $\xi^1$ are timelike and spacelike parameters, respectively. Thus, the string is characterized by its world sheet. The induced metric, $h_{ab}$, on the world sheet is given by

\begin{equation} \label{h}
h_{ab}=g_{\mu \nu} \frac{\partial x^\mu}{\partial \xi^a }\frac{\partial x^\nu}{\partial \xi^b }.
\end{equation}

According to the formalism used by Letelier, it can be associated to the string world sheet, a bivector $\Sigma^{\mu \nu}$, such that

\begin{equation}
\Sigma^{\mu \nu} = \epsilon^{ab} \frac{\partial x^\mu}{\partial \xi^a }\frac{\partial x^\nu}{\partial \xi^b },
\end{equation}

\noindent where $\epsilon^{ab}$ is the two-dimensional Levi-Civita symbol, with $ \epsilon^{01}=- \epsilon^{10} = 1$. 

Now, let us consider the energy-momentum tensor for a cloud of strings, characterized by a proper density $\rho$, which can be written as \cite{letelier1979clouds}

\begin{equation} \label{T1}
T^{\mu \nu}=\frac{\rho \Sigma^{\mu \beta} \Sigma^{\phantom{a}\nu}_\beta}{\sqrt{-h}},
\end{equation}

\noindent where $h=\frac{1}{2} \Sigma^{\mu \nu}\Sigma_{\mu \nu}$ is the determinant of the induced metric.

As the string cloud is spherically symmetric, the density, $\rho$, and the bivector $\Sigma^{\mu\nu}$, should be functions only of the radial coordinate. In this case, the non vanishing components of the bivector are $\Sigma^{01}$ and $\Sigma^{10}$, which are related by $\Sigma^{01}=-\Sigma^{10}$. Thus, $T^0_0=T^1_1=-\rho \Sigma^{01}$ and, using the relation \cite{letelier1979clouds}

\begin{equation}
\frac{d}{dr}[(r^2T^0_0)^\frac{1}{2}]=0,
\end{equation}

\noindent we find the following result

\begin{equation}
T^0_0=T^1_1=\frac{a}{r^2},
\end{equation}

\noindent where $a$ is an integration constant. 

Thus, let us solve the Einstein's equations for the source under consideration, which are written as

\begin{equation}
R_{\mu \nu}-\frac{1}{2}g_{\mu \nu}R =-\frac{\rho  \Sigma^{\phantom{a}\alpha}_\mu\Sigma^{\alpha \nu}}{\sqrt{-h}}.
\end{equation}

Assuming that the cloud of strings is uniformly distributed over a sphere, along the radial direction, the spacetime can be described by a general static spherically symmetric metric given by

\begin{equation} \label{met}
ds^2 = e^\nu dt^2- e^\lambda dr^2 - r^2 d \Omega^2,
\end{equation}

\noindent where $\nu = \nu(r)$ and $\lambda = \lambda(r)$. Now, taking into account that \cite{letelier1979clouds}

\begin{equation}
\Sigma^{01}=\frac{a}{\rho r^2} e^{-(\lambda+\nu)/2}.
\end{equation}

\noindent with the gauge invariant density, $\rho (-h)^{1/2}$, being $\rho (-h)^{1/2}=a/r^2$, in which $a$ should be a positive constant \cite{letelier1979clouds}. Thus, the Einstein's field equations can be written as

\begin{subeqnarray} \label{edo}
2 \nu''-\lambda' \nu'+4 \nu'/r+\nu'^2 = 0, \\
2 \lambda''-\lambda' \lambda'+4 \nu'/r+\lambda'^2 = 0,\\
e^{-\lambda}\left[1+\frac{r}{2}(\nu' - \lambda')\right]-1=-a,
\end{subeqnarray}

\noindent from which, we find that 

\begin{equation} \label{17}
\nu = -\lambda,
\end{equation}

\noindent and as a consequence 

\begin{equation} \label{12}
(e^{=\lambda}r)' =1-a.
\end{equation}

From Eq. (\ref{12}), we conclude that 

\begin{equation}
e^\nu = e^{-\lambda} = 1-a-\frac{2M}{r}.
\end{equation}

The solution obtained represents a static black hole surrounded by a spherically symmetric cloud of strings (Letelier 
spacetime) whose line element is

\begin{equation}
ds^2 = \left(1-a-\frac{2M}{r}\right)dt^2-\left(1-a-\frac{2M}{r}\right)^{-1} dr^2 - r^2 d \Omega^2,
\end{equation}
where $M$ is the mass parameter.

The modification in the metric as compared with the Schwarzschild one is due exclusively to the parameter $a$, which codifies the presence of the cloud of strings. The case $a=0$, corresponds to the Schwarzschild solution. For $a \neq 0$, the solution is formally analogous to the one associated with the global monopole \cite{barriola1989gravitational}. 

It is worth calling attention to the fact that, when $M=0$, there is only a cloud of strings. If we consider the string as straight and infinitely thin, its gravitational mass vanishes due to the fact that the gravitational effect produced by tension cancels the effect generated by its mass\cite{vilenkin1981cosmic,gott1985gravitational,hiscock1985exact}. This means that locally there is no gravity and the gravitational effects, in this case, arise in connection with the global ones associated with a deficit planar angle. With respect to the cloud of strings, the gravitational effects are also of global origin, but now connected with a solid deficit angle analogously to what happens in the case of a global monopole \cite{barriola1989gravitational}.

The Letelier spacetime has a horizon of radius

\begin{equation} \label{r1}
r_h = \frac{2M}{1-a}, \qquad a \neq 1.
\end{equation}

\noindent which tends to the Schwarzschild horizon, when $a \rightarrow 0$. Otherwise, when $a \rightarrow 1$, the radius of the event horizon tends to infinity. 

Equation (\ref{r1}) tells us that the presence of the cloud of strings enlarges the radius of the horizon event as compared with the Schwarzschild radius by a factor $(1-a)^{-1}$.  

\subsection{Kiselev spacetime}

Recently, it was obtained the solution corresponding to a black hole immersed in quintessence, whose energy-momentum tensor is given by \cite{kiselev2003quintessence}

\begin{subeqnarray}
T^t_{\phantom{t}t} = T^r_{\phantom{r}r} = \rho_q, \\
T^\theta_{\phantom{t}\theta} = T^\phi_{\phantom{t}\phi} = -\frac{1}{2} \rho_q (3 \omega_q+1),
\end{subeqnarray}

\noindent where the pressure and density of quintessence, $p_q$ and $\rho_q$, respectively, are related by the equation of state $p_q = \omega_q \rho_q$, with $\omega_q$ being the quintessential state parameter.

If we consider this mass-energy content and adopt the static spherically symmetric metric background given by Eq. (\ref{met}), the Einstein's field equations are given by \cite{kiselev2003quintessence}

\begin{subeqnarray} \label{edo2}
-e^{-\lambda}\left(\frac{1}{r^2}+\frac{\lambda'}{r} \right)+\frac{1}{r^2} = 2 \rho_q, \\
-e^{-\lambda}\left(\frac{1}{r^2}+\frac{\nu'}{r} \right)+\frac{1}{r^2} = 2 \rho_q,\\
\frac{1}{2}e^{-\lambda}\left(\nu''+\frac{v'^2}{2}+\frac{\nu' - \lambda'}{r}-\frac{\nu' \lambda'}{2}\right)= \rho_q(3 \omega_q+1).
\end{subeqnarray}

In this case, we also obtain the same relation given by Eq. (\ref{17}) \cite{kiselev2003quintessence}. Thus, making the substitution $\lambda = -ln (1+f(r))$, where $f(r)$ is a function to be determined, into Eq. (\ref{edo2}) and combining the resulting equations, we find the following differential equation for the function $f(r)$ \cite{kiselev2003quintessence}

\begin{equation} \label{fr}
r^2f''+3(\omega_q+1)r f'+(3 \omega_q +1) f=0,
\end{equation}

\noindent which is a Euler homogeneous differential equation whose general solution can be written as\cite{kiselev2003quintessence}

\begin{equation} \label{sol1}
f(r) = -\frac{2M}{r}-\frac{\alpha}{r^{3\omega_q+1}},
\end{equation}

\noindent with $\alpha$ and $M$ being integration constants. With this solution, we can write explicitly for the spacetime under consideration, which is written as

\begin{equation}
ds^2 = \left(1 -\frac{2 M}{r} - \frac{\alpha}{r^{3\omega_q+1}}\right)dt^2 \nonumber-\left(1 -\frac{2 M}{r} - \frac{\alpha}{r^{3\omega_q+1}}\right)^{-1}dr^2 - r^2 d \Omega^2.
\end{equation}

This solution corresponds to a static black hole surrounded by quintessence, whose quintessence density $\rho_q$ is given by

\begin{eqnarray}
\rho_q = - \frac{\alpha}{2} \frac{3 \omega_q}{r^{3 ( \omega_q +1)}}.
\end{eqnarray}

In order to get the scenario of accelerated expansion, it is necessary to impose that $- 1<\omega_{q}< - 1/3$. As to $q$, it is a positive parameter and such that, when $q = 0$, we recover the Schwarzschild solution. Thus, we conclude that the quintessence parameter $\alpha$ should be positive \cite{kiselev2003quintessence}.

\section{Letelier spacetime with quintessence}
\label{sec3}

Now, let us consider a static spherically symmetric black hole surrounded by a cloud of strings and quintessence. Supposing that the quintessence and the cloud of strings do not interact with each other, we can assume that the energy-momentum tensor is obtained by a linear superposition of the energy-momentum tensors corresponding to each one of the cases, discussed in section \ref{sec2}. Then, we get

\begin{subeqnarray}
T^t_{\phantom{t}t} = T^r_{\phantom{r}r} = \rho_q+\frac{a}{r^2}, \\
T^\theta_{\phantom{t}\theta} = T^\phi_{\phantom{t}\phi} = -\frac{1}{2} \rho_q (3 \omega_q+1).
\end{subeqnarray}

Considering, once again, the general static spherically symmetric line element given by Eq. (\ref{met}), the Einstein's equations can be written as

\begin{eqnarray} \label{edo3}
-e^{-\lambda}\left(\frac{1}{r^2}+\frac{\lambda'}{r} \right)+\frac{1}{r^2} = 2 \rho_q+\frac{2a}{r^2}, \label{edo3a} \\
-e^{-\lambda}\left(\frac{1}{r^2}+\frac{\nu'}{r} \right)+\frac{1}{r^2} = 2 \rho_q+\frac{2a}{r^2}, \label{edo3b}\\
\frac{1}{2}e^{-\lambda}\left(\nu''+\frac{v'^2}{2}+\frac{\nu' - \lambda'}{r}-\frac{\nu' \lambda'}{2}\right)= \rho_q(3 \omega_q+1). \label{edo3c}
\end{eqnarray}

Combining Eqs. (\ref{edo3a}) and (\ref{edo3b}), we find that $\lambda=-\nu$, which is the same relation of the previous cases, as should be expected. Let us once more take $\lambda = -ln(1+f(r))$. Thus, substituting this relation into Eqs. (\ref{edo3a})-(\ref{edo3c}), and combining resulting equations appropriately, we find the following differential equation for $f(r)$   

\begin{equation} \label{fr2}
r^2f''+3(\omega_q+1)r f'+(3 \omega_q +1) f+a(3 \omega_q+1)=0.
\end{equation}

Note that Eq. (\ref{fr2}) is a non-homogeneous Euler differential equation, whose solution can be written as

\begin{equation} 
f = f_h+f_p,
\end{equation}

\noindent where $f_h$ is the solution of the corresponding homogeneous equation and $f_p$ is a particular solution related to the non-homogeneous part of the differential equation. It is straightful conclude that the homogeneous part of Eq. (\ref{fr2}) corresponds to Eq. (\ref{fr}). Then, $f_h$ is given by Eq. (\ref{sol1}). Moreover, the particular solution is given by $f_p=-a$. Thus, we get

\begin{equation} \label{sol2}
f(r) = -\frac{2M}{r}-\frac{\alpha}{r^{3\omega_q+1}}-a
\end{equation}

Substituting Eq. (\ref{sol2}) into Eq. (\ref{met}), we obtain the spacetime metric concerning the black hole surrounded by a cloud of strings and quintessence, which is given by

\begin{equation} \label{met2}
ds^2 = g(r)dt^2 \nonumber- g(r)^{-1}dr^2 - r^2 d \Omega^2.
\end{equation}
with 
\begin{equation}
g(r) = 1-a -\frac{2 M}{r} - \frac{\alpha}{r^{3\omega_q+1}}.
\end{equation}

\noindent  This metric corresponds to the Letelier spacetime with quintessence. 
Observe that the cloud of strings introduces in the metric the constant term $a$ which is similar to the one that appears in the spacetime of the Schwarzschild black hole with a global monopole, already considered in the literature \cite{rodrigue2018thermodynamics}.

\section{Spacetime singularities and horizons}
\label{sec4}

In what follows, let us consider the singularities and structures of the horizons associated with the Letelier spacetime with quintessence, given by the line element shown in Eq. (\ref{met2}).

Singularities are points in which the line element degenerates. In the case under consideration, the spacetime singularities can be found from the equation

\begin{equation} \label{g}
g(r) = 1-a-\frac{2M}{r}-\frac{\alpha}{r^{3\omega_q+1}} = 0.
\end{equation}

First of all, note that, at the point $r=0$, the metric of the Letelier spacetime with quintessence diverges,
which is a behavior similar to what occurs in the case of the Schwarzschild spacetime. Taking into account Eq. (\ref{met2}), 
the Riemann curvature scalar is given by 

\begin{equation}
R=\frac{2 a }{r^2}+\frac{3 \alpha \omega_q (1-3 \omega_q)}{r^{3(\omega_q+1)}},
\end{equation}

\noindent which also diverges at $r=0$, but differently from the Schwarzschild spacetime, is not equal to zero for others values of $r$.

The other singularities that come from Eq. (\ref{g}) can be removable under an appropriate coordinate transformation. Then, they are called removable singularities and they define the metric horizons. 

\begin{figure}[!htb]
    \centering
    \includegraphics[scale=0.75]{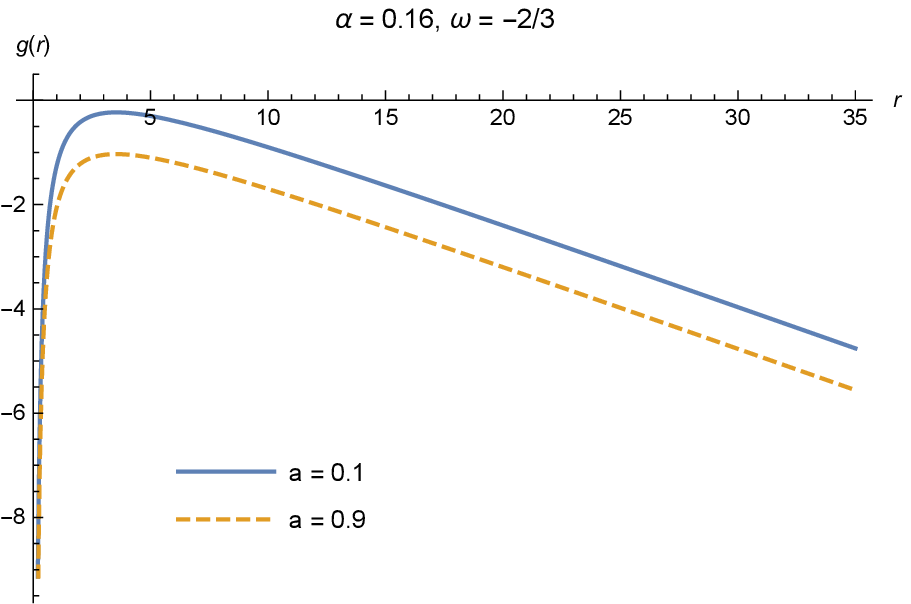}
    \includegraphics[scale=0.75]{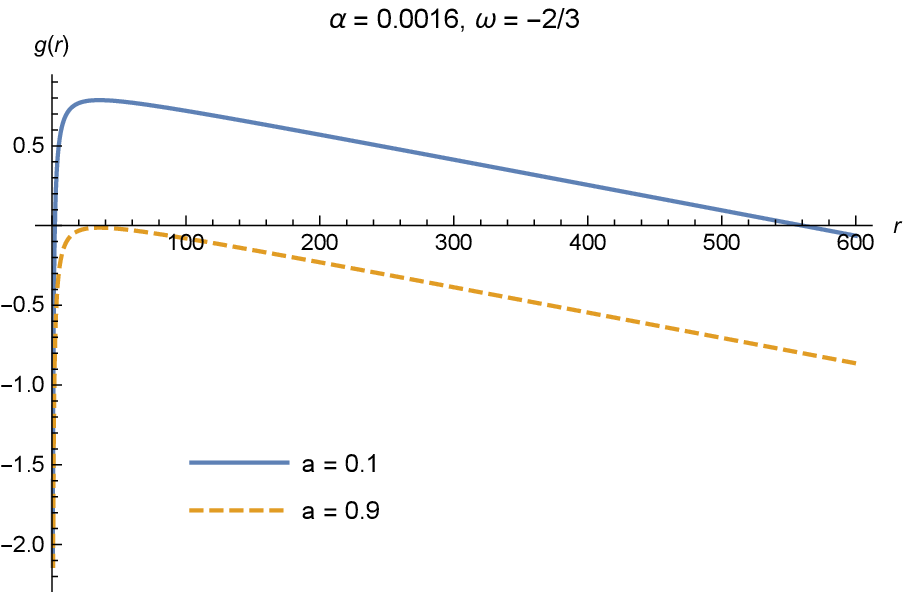}
    \includegraphics[scale=0.75]{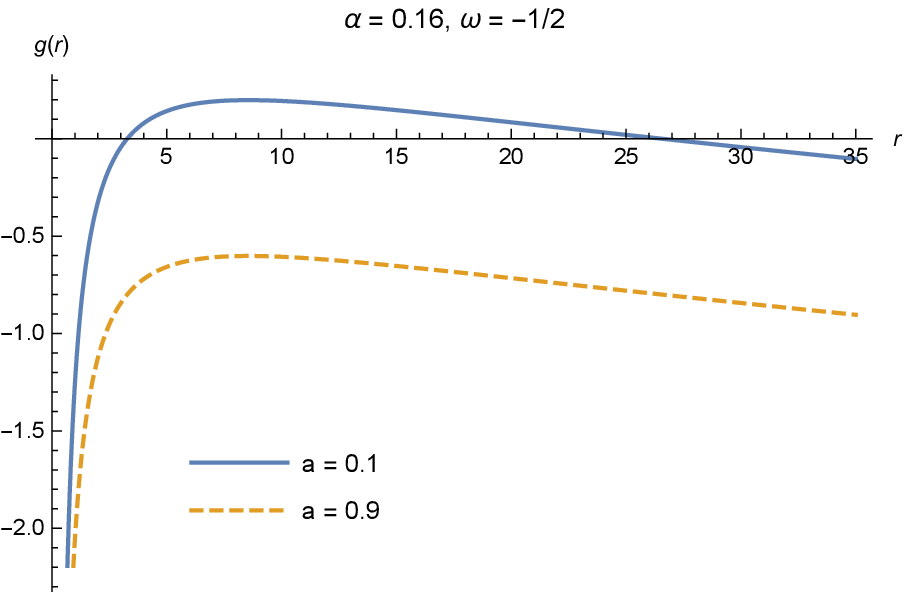}
    \includegraphics[scale=0.75]{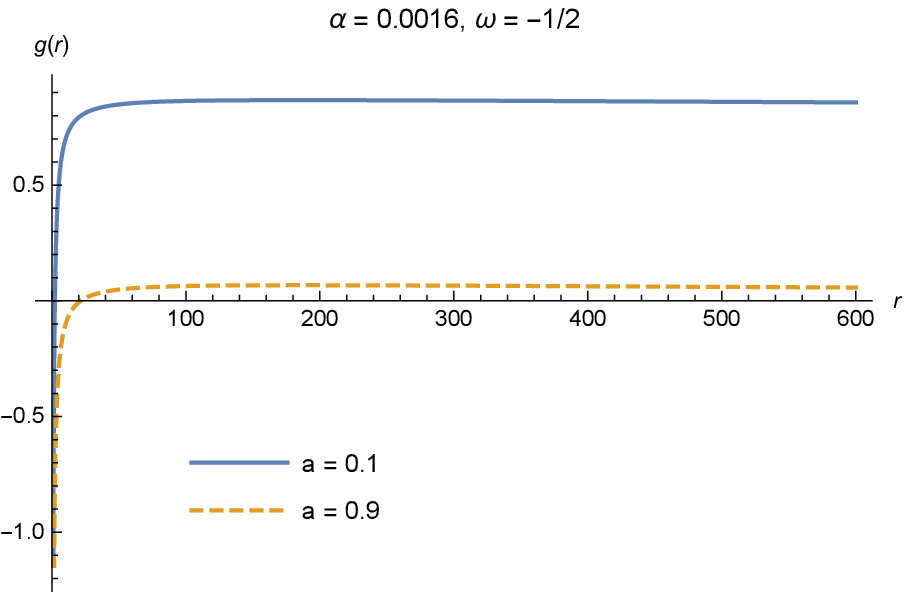}
    \caption{The function $g(r)$ for different values of the parameters $a$, $\alpha$ and $\omega_q$. }
    \label{Fig1}
  \end{figure}

In Fig. \ref{Fig1}, we represent the function $g(r)$ for different values of $a$, $\alpha$ and $\omega_q$. The horizons of the black hole are defined by the roots of the function $g(r)$.

Observe that, depending on the values of the parameters of the mass-energy configuration we are dealing with, the number of horizons of the black hole can assume values from zero to two. When the studied black hole has two horizons, the internal one is the event (black hole) horizon and the external one is the cosmological horizon, which is related to the quintessential fluid.

If we take the particular case $\omega_q = -2/3$, Eq. (\ref{g}) is reduced to

\begin{equation}
\alpha r^2 - (1-a) r + 2 M = 0.
\end{equation}

This equation define the horizons. Its solutions are given by

\begin{subeqnarray}
r_- = \frac{1-a - \sqrt{(1-a)^2 - 8 M \alpha}}{2 \alpha},\\
r_+ = \frac{1-a + \sqrt{(1-a)^2 - 8 M \alpha}}{2 \alpha}.
\end{subeqnarray}

Note that, if $(1-a)^2 < 8 M \alpha$, the black hole will have no horizon. In this case, it will be called a naked singularity. If $(1-a)^2 = 8 M \alpha$, the black hole has just one horizon. Finally, if $(1-a)^2 > 8 M \alpha$, the black hole has two horizons: the event horizon, $r_-$, and a cosmological horizon, $r_+$. Therefore, the number of horizons depends on the relation between the parameters which codify the presence of the cloud of strings and of the quintessence.

\section{Black hole thermodynamics}
\label{sec5}

In this section, we study the thermodynamics of the black hole considered by examining the behaviors of the mass, temperature, and heat capacity as a function of the entropy.

At any horizon with radius $r_h$, we have that $g(r_{h})=0$. This codition gives us the mass

\begin{equation} \label{M}
M = \frac{1-a}{2} r_h -\frac{\alpha}{2 r_h^{3 \omega_q}},
\end{equation}
which is written in terms of the parameters that codify the presence of the cloud of string as well as of the quintessence. Note that 
when $a=0$ and $\alpha=0$, we recover the mass of the Schwarzschild black holein terms of the horizon radius. Therefore, expression given 
by Eq. (\ref{M}) repreents the mass(energy) of the black hole, appropriately modified by the presence of the cloud of strings and the quintessence. 

The horizon area can be calculated by

\begin{equation}
A = \int \sqrt{-g} d \theta d \phi = 4 \pi r_h^2.
\end{equation}

On the other hand, the area law tells us that the black hole entropy is given by \cite{bekenstein1973black,rodrigue2018thermodynamics}
\begin{equation}
S = \frac{A}{4} = \pi r_h^2.
\end{equation}

\begin{figure}[!htb]
    \centering
    \includegraphics[scale=0.75]{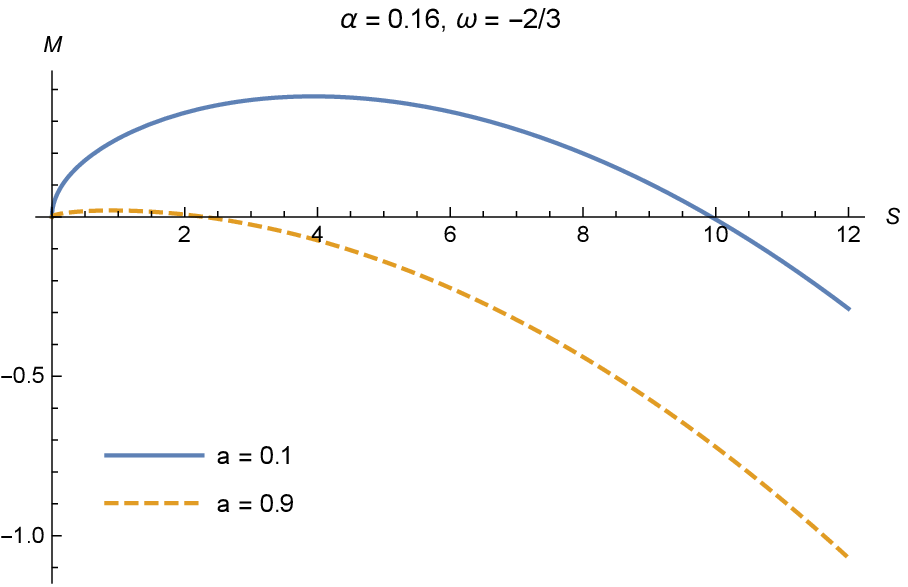}
    \includegraphics[scale=0.75]{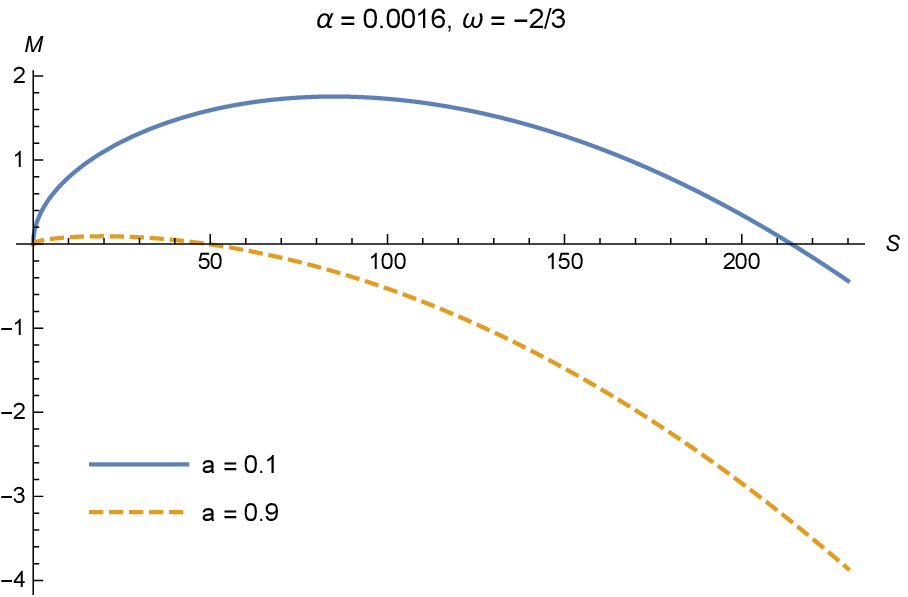}
    \includegraphics[scale=0.75]{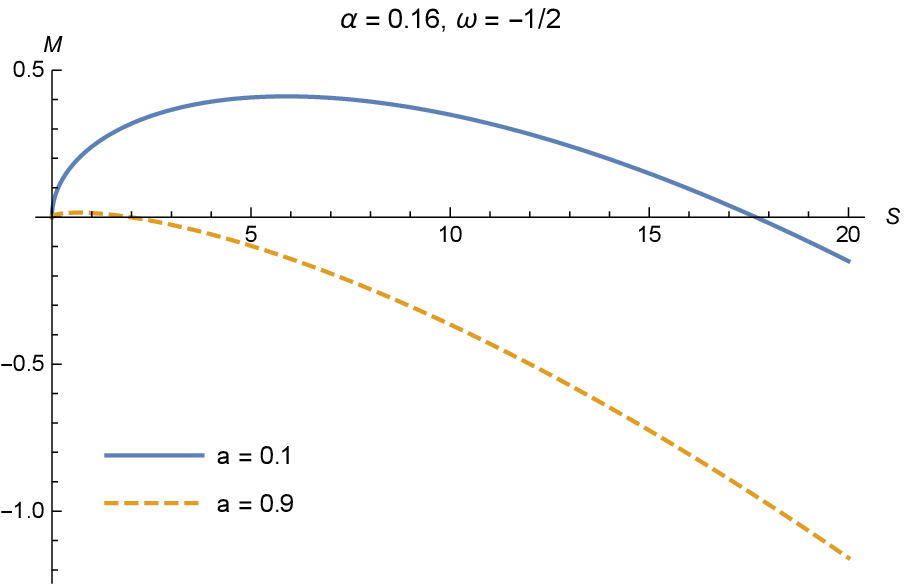}
    \includegraphics[scale=0.75]{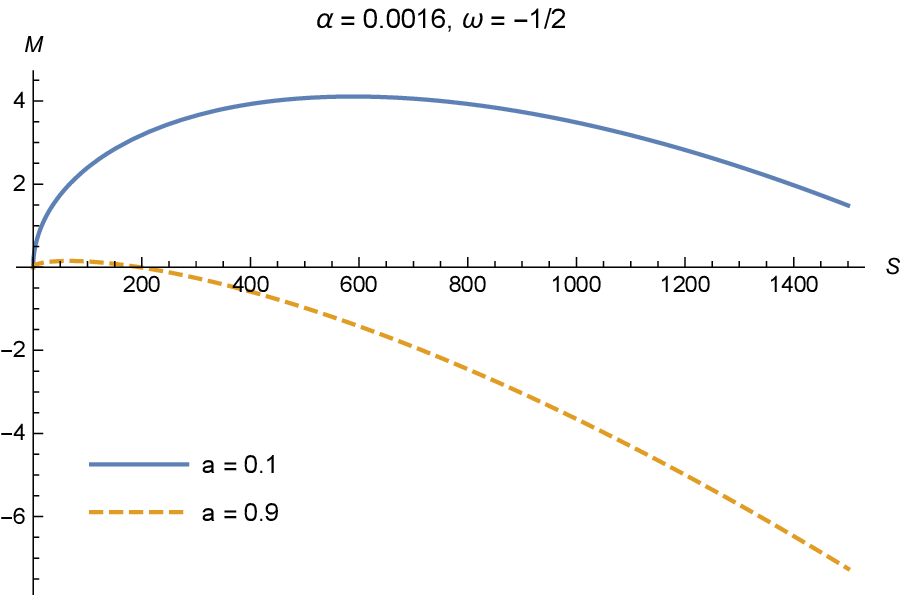}
    \caption{Mass parameter as a function of the entropy for different values of $a$, $\alpha$ and $\omega_q$. }
    \label{Fig2}
  \end{figure}

Thus, we can write the mass parameter as a function of the black hole entropy, which can be written as

\begin{equation}
M = \frac{1-a}{2}\sqrt{\frac{S}{\pi}} -\frac{\alpha}{2}\left( \frac{\pi}{S} \right)^{3 \omega_q},
\end{equation}

\noindent whose behavior, as a function of entropy, is shown in Fig. \ref{Fig2}, for different values of  $a$, $\alpha$ and $\omega_q$.

\begin{figure}[!htb]
    \centering
    \includegraphics[scale=0.75]{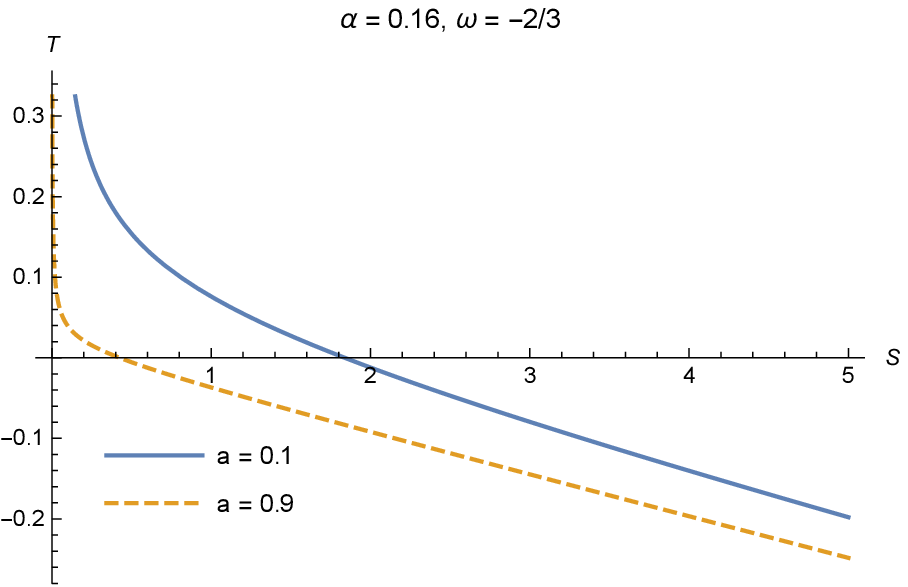}
    \includegraphics[scale=0.75]{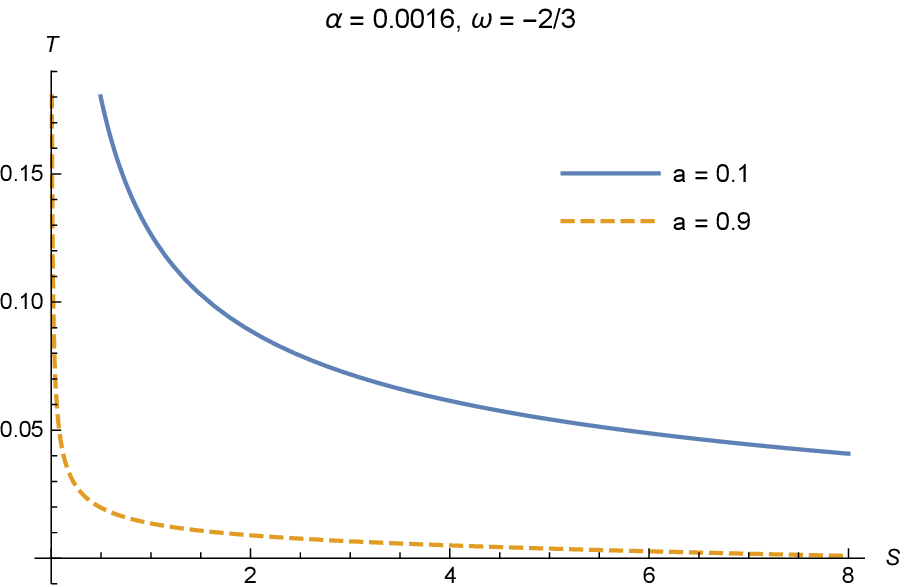}
    \includegraphics[scale=0.75]{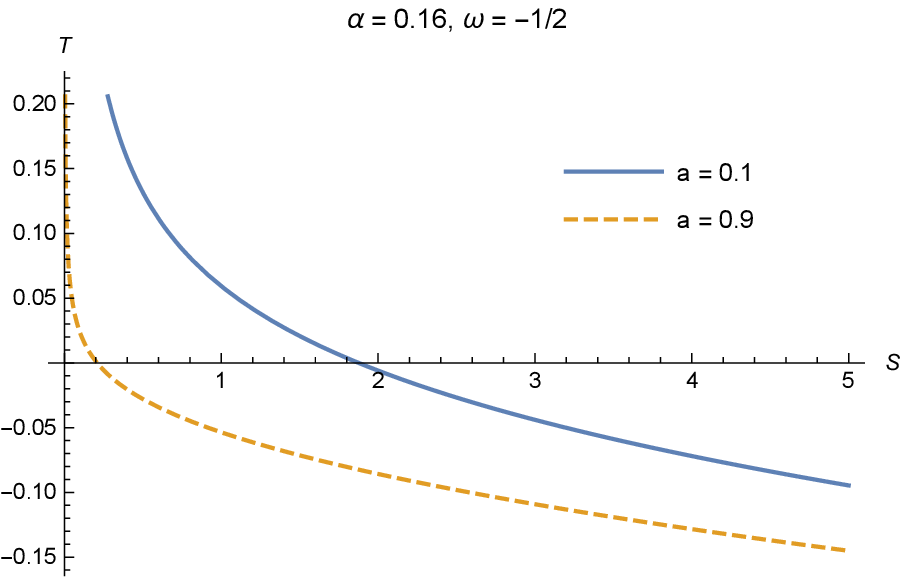}
    \includegraphics[scale=0.75]{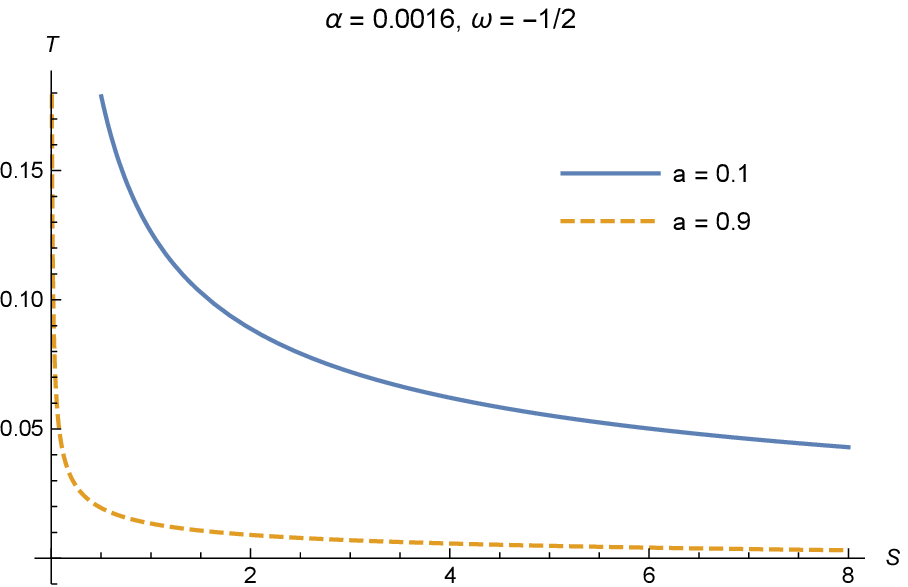}
    \caption{Hawking temperature as a function of the entropy for different values of $a$, $\alpha$ and $\omega_q$.}
    \label{Fig3}
  \end{figure}

The Hawking temperature of the black hole horizon is obtained by

\begin{equation} \label{temp}
T = \frac{\partial M}{\partial S} = \frac{1-a}{4 \sqrt{\pi S}} +\frac{3 \alpha \omega_q}{2 S}\left( \frac{\pi}{S} \right)^{3 \omega_q},
\end{equation}
whose behavior in terms of the entropy is shown in Fig. \ref{Fig3}, for different values of the cloud of strings and quintessence parameters.

\begin{figure}[!htb]
    \centering
    \includegraphics[scale=0.75]{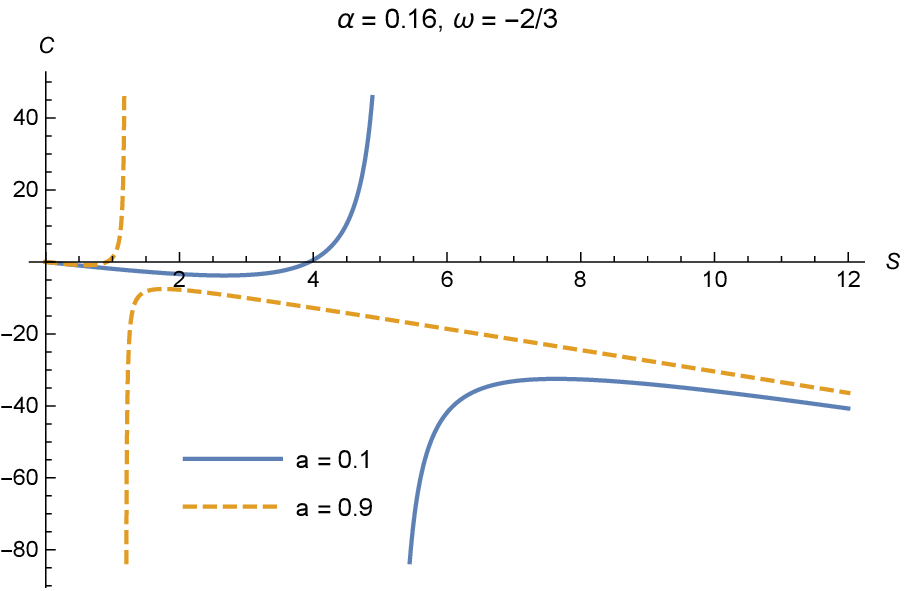}
    \includegraphics[scale=0.75]{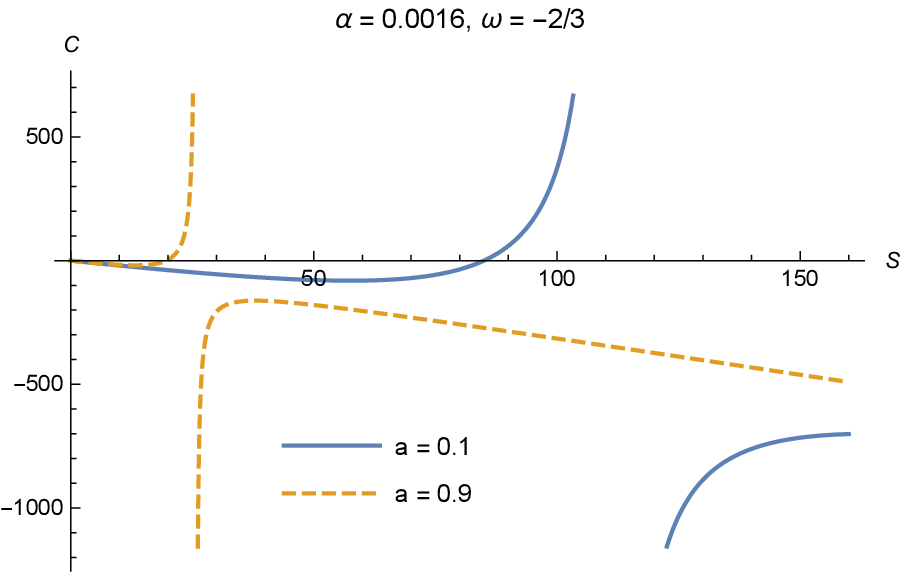}
    \includegraphics[scale=0.75]{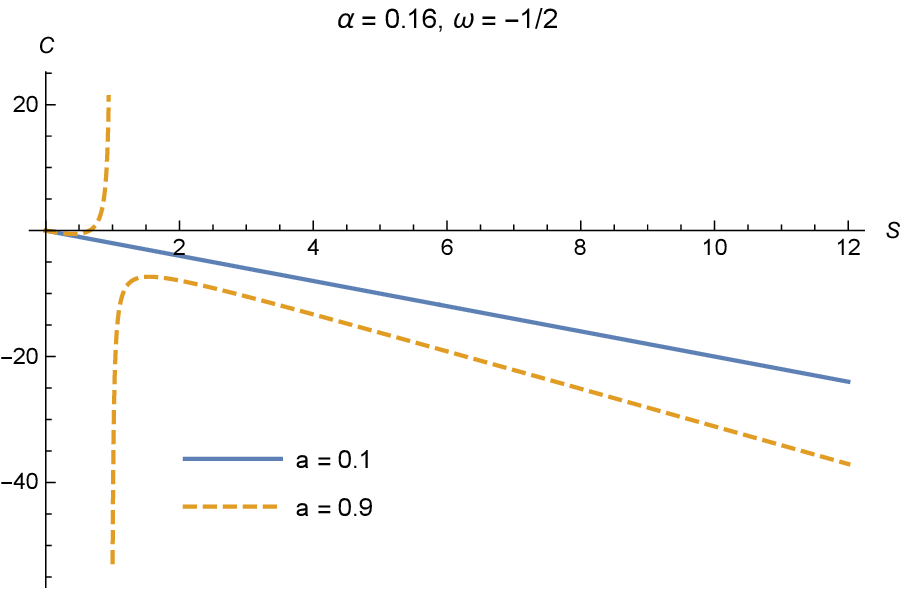}
    \includegraphics[scale=0.75]{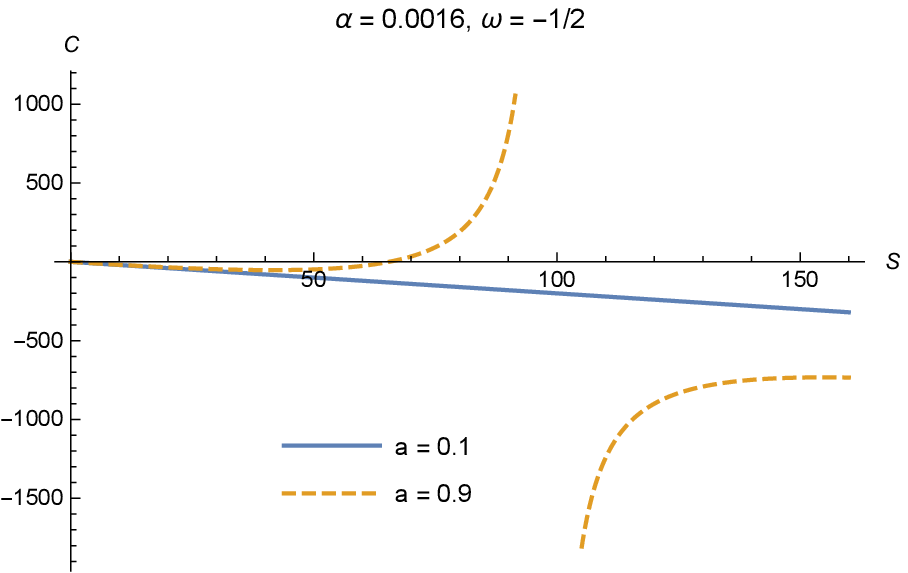}
    \caption{Heat capacity as a function of the entropy for different values of $a$, $\alpha$ and $\omega_q$.}
    \label{Fig4}
  \end{figure}

With the entropy as a function of the temperature, obtained from Eq. (\ref{temp}), we can get the expression for the heat capacity associated with the black hole, which is written as

\begin{equation} \label{cap}
C = T\frac{\partial S}{\partial T} = -\frac{2(1-a) S^{3 \omega_q+3/2}+ 12 \alpha \omega_q \pi^{3 \omega_q+1/2} S}{(1-a)S^{3 \omega_q +1/2}+12 \alpha \omega_q(3\omega_q+1)\pi^{3 \omega_q+1/2}},
\end{equation}
whose behavior is shown in Fig. \ref{Fig4}, as a function of the black hole entropy, for different values of the cloud of string and quintessence parameters.

In the absence of the cloud of strings and the quintessence, Eq. (\ref{cap}) is reduced to $C=-2S$. This is an expected result, because 
in this particular situation, the black hole we are considering reduces to the Schwarzschild black hole.

Otherwise, when we take into account the effect of the cloud of strings and the quintessence, there are values of the entropy for which the heat capacity acquires positive values. It means that the black hole can be thermodynamically unstable or stable, depending on its entropy, and, as can be seen in Fig. (\ref{Fig3}), there are phase transitions. It is important to highlight that the cloud of string parameter plays a role in the behavior of the heat capacity, modifying the phase transition point.

\section{Hawking radiation}
\label{sec6}

Now, we will consider the phenomenon pointed out by Hawking \cite{hawking1975particle} concerning the radiation of particles by a black hole, and emphasize the role played by the combination of the cloud of strings and quintessence. Using tortoise and advanced Eddington-Finkelstein coordinates

\begin{equation}
\frac{dr_*}{dr}=g(r)^{-1},
\end{equation}

\begin{equation}
t=T-r_*,
\end{equation}

\noindent the metric given by Eq. (\ref{met2}) turns into \cite{rodrigue2018thermodynamics}:

\begin{equation} 
ds^2 = -g(r) dT^2 + 2dT dr + r^2 (d \theta^2+ d \phi^2).
\end{equation}

As we can verify,  $r=r_+$ is not a singularity of this spacetime \cite{li2017hawking}. The null radial geodesic is obtained from

\begin{equation}
\frac{dr}{dT} = \frac{g(r)}{2}.
\end{equation}

Therefore, the imaginary part of the action corresponding to the particle which crosses the horizon is given by:

\begin{equation}
Im Z = Im \int^{T_f}_{T_i} L dT = Im \int^{r_f}_{r_i} p_r dr =  Im \int^{r_f}_{r_i} \int^{p_r}_{0} dp_r dr,
\end{equation}

\noindent where $p_r$ is the canonical momentum associated with the coordinate $r$. Thus, using the relation $\dot{r} = \frac{dH}{dp_r} \big|_r$ and the fact that $(dH_r) = dM$, we find the following result:

\begin{equation}
Im Z =  Im \int^{M_f}_{M_i} \int^{r_f}_{r_i}\frac{dr}{\dot{r}}dM =  Im \int^{M_f}_{M_i} \int^{r_f}_{r_i}\frac{2}{g(r)}drdM,
\end{equation}

\noindent where $M_i = M$ is the original mass of the black hole and $M_f = M-\omega_0$ is the mass after the emission of a particle with energy $\omega_0$. Note that the integrand has a pole at $r=r_+$, and thus we can calculate the contour integral around that pole as \cite{li2017hawking}

\begin{equation} \label{49}
ImZ = -2 \pi \int^{M_f}_{M_i} \frac{1}{g'(r_+)}dM = -2 \pi \int^{r_{+f}}_{r_{+i}} \frac{1}{g'(r_+)}\frac{dM}{dr_+}dr_+.
\end{equation}

From Eqs. (\ref{g}) and (\ref{M}), we find

\begin{equation}
ImZ = -\frac{\pi}{2}(r_{+f}^2-r_{+i}^2)= -\frac{\Delta S}{2}.
\end{equation}

Using a result from the WKB method, which tells us that the probability that a particle emitted by the black hole experience a tunneling is given by \cite{li2017hawking,srinivasan1999particle,parikh2000hawking}

\begin{equation}
\Gamma \sim exp(-2 Im Z),
\end{equation}

\noindent we find that

\begin{eqnarray}
ln \Gamma \sim  - \Delta S
\end{eqnarray}

\noindent or

\begin{equation}
\Gamma \sim e^{\Delta S},
\end{equation}

\noindent which means that this probability is related to the change of the entropy of the black hole. Thus it depends on the event horizon which is influenced by the parameters associated with the presence of the cloud of strings as well as of the quintessence.

From Eq. (\ref{49}), it is possible to write the black hole decay rate as \cite{srinivasan1999particle,parikh2000hawking}

\begin{equation}
\Gamma \sim e^{-\beta \omega_0},
\end{equation}

\noindent where

\begin{equation}
\beta= \frac{1}{T}
\end{equation}

\noindent is the Boltzmann factor. Now, if we write the cosmological horizon radius $r_+$ as \cite{rodrigue2018thermodynamics}

\begin{equation}
r_+=r_0+\delta,
\end{equation}

\noindent where

\begin{equation}
r_0=\frac{2M}{1-a}
\end{equation}

\noindent is the horizon coordinate in the absence of quintessence, in the first order of approximation, the correction of the horizon radius is expressed as 
\begin{equation}
\delta=\frac{\alpha}{(1-a) r_0^{3 \omega_q}}.
\end{equation}

Thus, the Boltzmann factor is given by \cite{rodrigue2018thermodynamics}

\begin{equation}
\beta_+ =\frac{4 \pi}{1-a}\left(\frac{2M}{1-a}+\frac{\alpha}{(1-a)r_0^{3 \omega_q}}\right).
\end{equation}

\begin{figure}[!htb]
    \centering
    \includegraphics[scale=0.75]{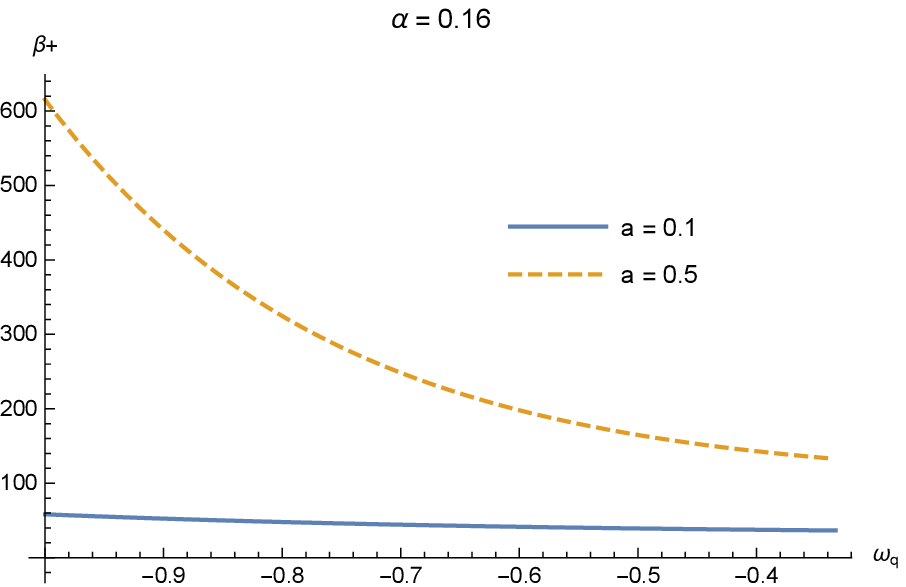}
    \includegraphics[scale=0.75]{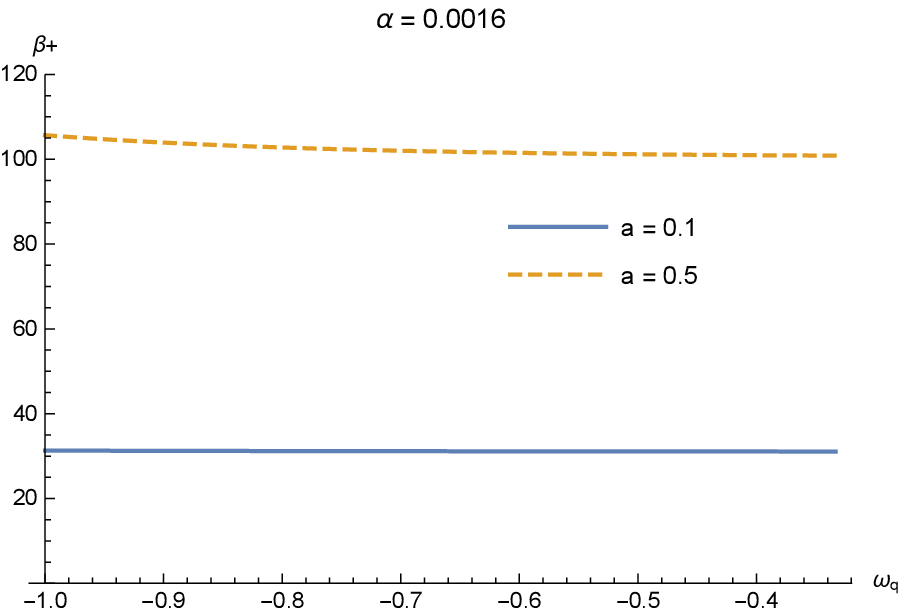}
    \caption{Boltzmann factor, $\beta_+$, as a function of  $\omega_q$ for different values of $a$ and $\alpha$.}
    \label{Fig5}
  \end{figure}

\begin{figure}[!htb]
    \centering
    \includegraphics[scale=0.75]{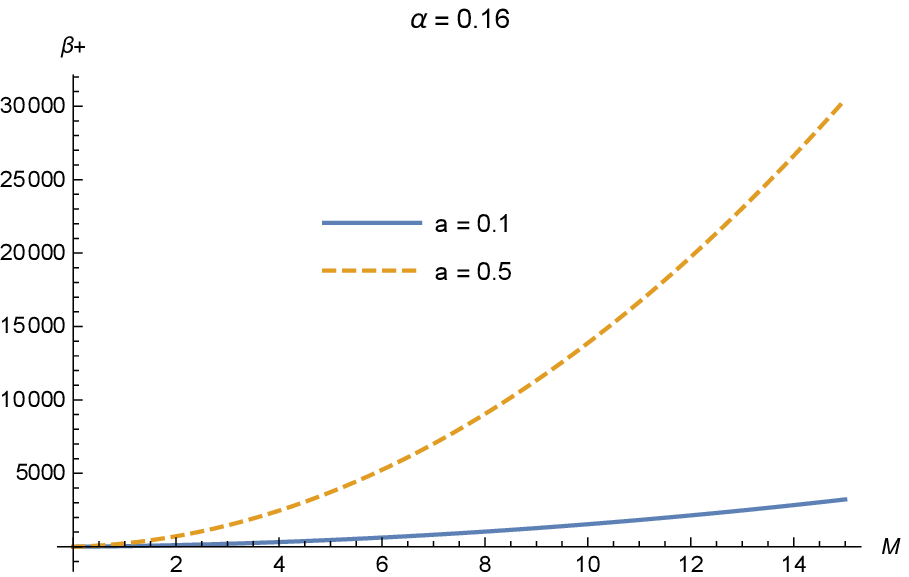}
    \includegraphics[scale=0.75]{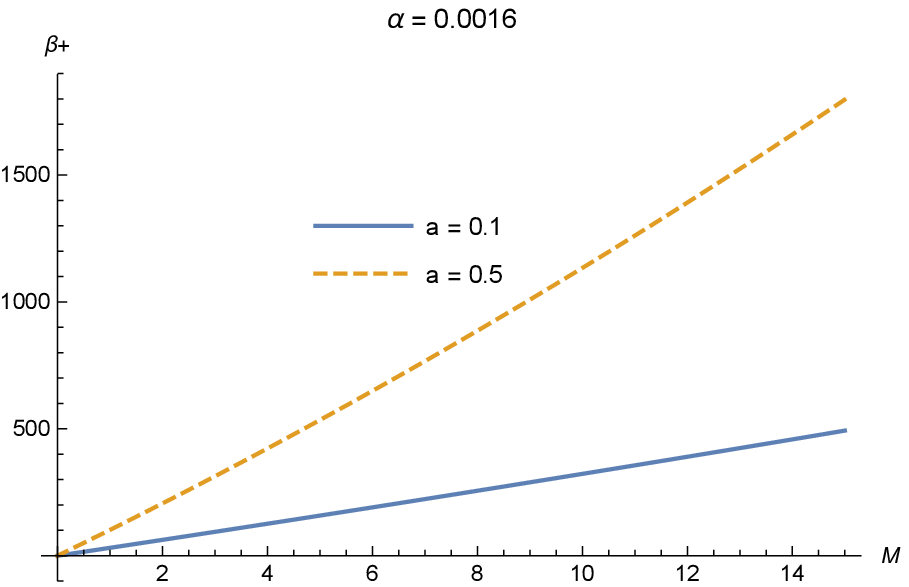}
    \caption{Boltzmann factor, $\beta_+$, as a function of  $M$ for different values of $a$ and $\alpha$.}
    \label{Fig6}
  \end{figure}

In Figs. \ref{Fig5} and \ref{Fig6}, is represented the behavior of the Boltzmann factor near the cosmological horizon as a function of $\omega_q$ and $M$, respectively.

\begin{figure}
    \centering
    \includegraphics[scale=0.75]{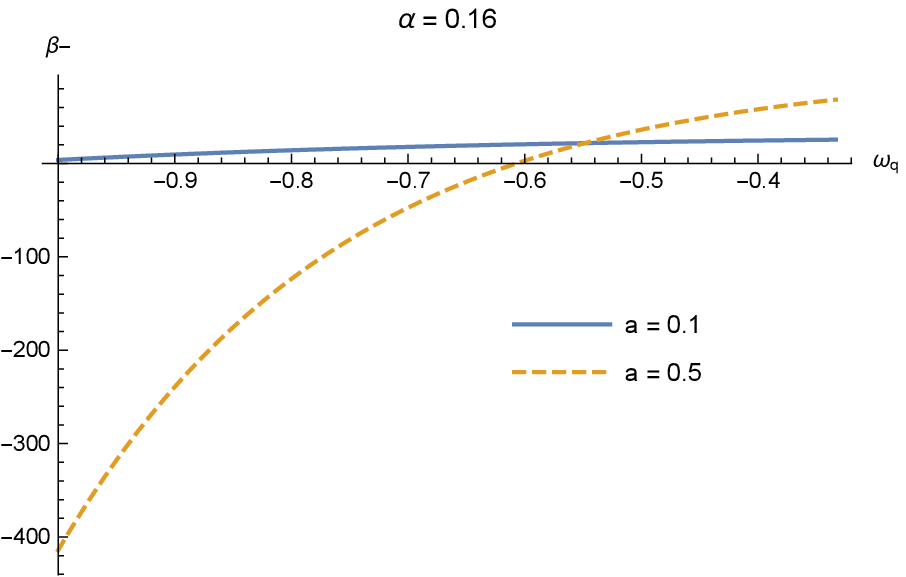}
    \includegraphics[scale=0.75]{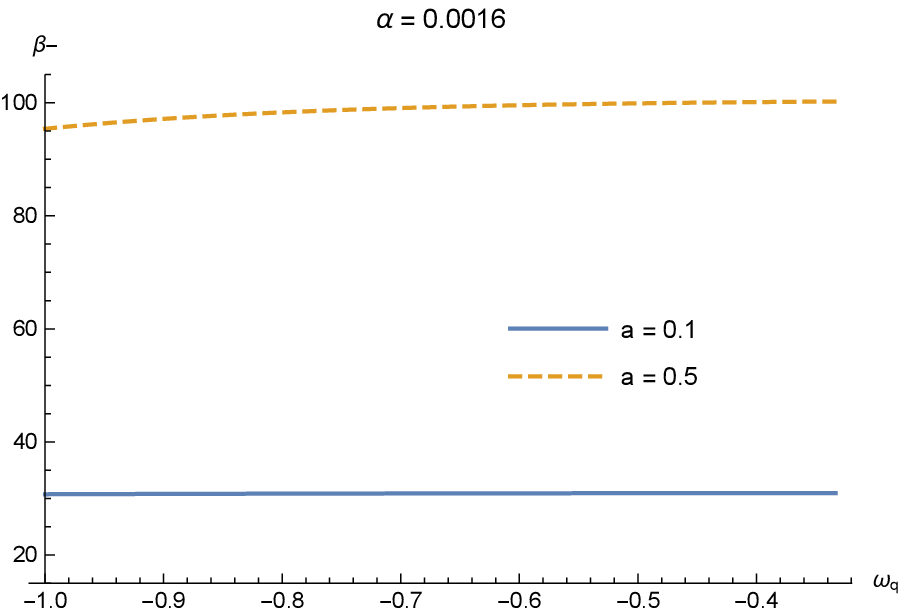}
    \caption{Boltzmann factor, $\beta_-$, as a function of  $\omega_q$ for different values of $a$ and $\alpha$.}
    \label{Fig7}
  \end{figure}

\begin{figure}
    \centering
    \includegraphics[scale=0.75]{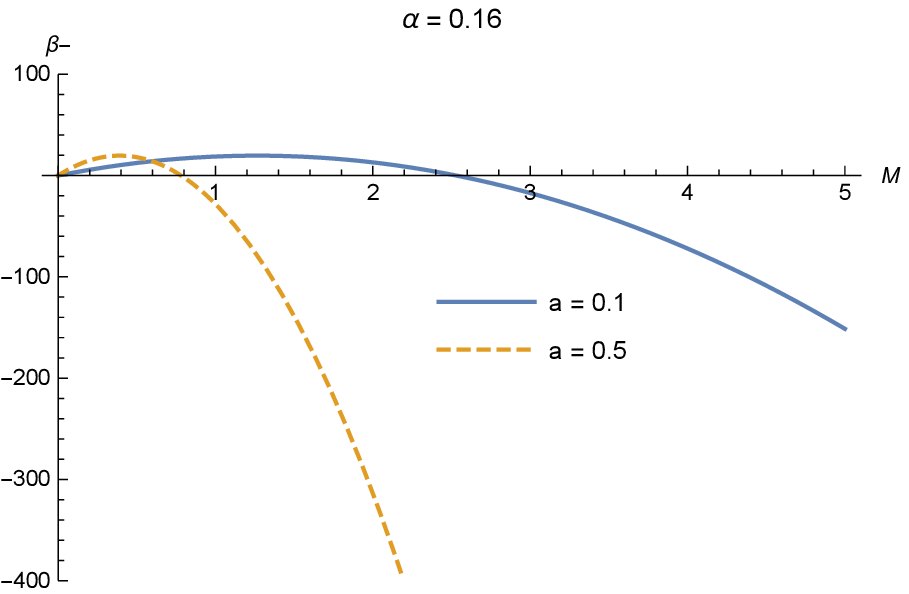}
    \includegraphics[scale=0.75]{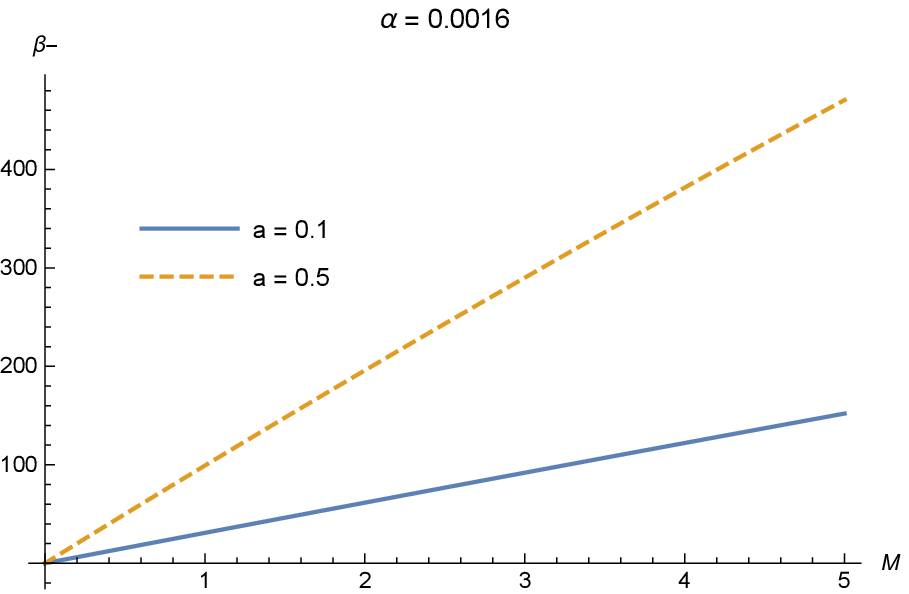}
    \caption{Boltzmann factor, $\beta_-$, as a function of  $M$ for different values of $a$ and $\alpha$.}
    \label{Fig8}
  \end{figure}

On the other hand, the event horizon $r_-$ can be written as

\begin{equation}
r_-=r_0-\delta,
\end{equation}

\noindent and, thus, the Boltzmann factor is given by

\begin{equation}
\beta_-=\frac{4 \pi}{1-a}\left(\frac{2M}{1-a}-\frac{\alpha}{(1-a)r_0^{3 \omega_q}}\right),
\end{equation}

\noindent whose behavior is represented, respectively, as a function of $\omega_q$ and $M$ in Figs. \ref{Fig7} and \ref{Fig8}.

Note that, the Boltzmann factor due to the Hawking radiation near the black hole horizon can assume negative values depending on the parameters that codify the presence of the cloud of strings and quintessence and of the mass of the black hole.

\section{Concluding remarks}
\label{sec7}

The solution of the Einstein equations corresponding to a black hole with a cloud of strings and quintessence (Letelier spacetime with quintessence) is similar to the solution which corresponds to a black hole with global monopole surrounded by quintessence. The role played by the parameter which codifies the presence of the cloud of strings is similar to the one associated to the deficit solid angle which depends on the scale of the symmetry breaking of a global monopole.

With respect to the number of horizons, it is strictly connected with the relation between the values of the parameter which codifies the presence of the cloud of strings and the parameter of quintessence.

In the study of the thermodynamics, the energy, temperature and heat capacity were obtained in terms of the entropy for different values of the parameters involved. The energy and temperature decrease when $a$ and $\alpha$ increase. Differently from the Schwarzschild black hole, which is thermodynamically unstable, the black hole with a cloud of strings and quintessence presents regions in which it is thermodynamically stable, depending on the values of the parameters $a$ and $\alpha$, for fixed values of $\omega_q$.
 
Finally, in relation to Hawking radiation, we analyzed the Boltzmann factor whose behaviors in terms of the quintessential state parameter, the parameters that codify the presence of the cloud of strings and quintessence and of the mass parameter are shown in 
Figs. \ref{Fig5} and \ref{Fig6}, in the case where the radiation is emmitted from the cosmological horizon, and in Figs. \ref{Fig7} and 
\ref{Fig8}, when the emission arises from the event horizon.

In summary, we conclude that all thermodynamics quantities are depending on the parameters involved in the construction of the solution corresponding to the black hole surrounded by a cloud of strings and quintessence, namely, the parameter which codifies the presence of the cloud of strings, the quintessential parameter, and the quintessential state parameter.

\begin{acknowledgements}
M. M. D. C. was supported by CAPES Fellowship. V. B. Bezerra is partially supported by Conselho Nacional de Desenvolvimento Cient\'ifico e Tecnol\'ogico (CNPq) through the research Project nr. 305835/2016-5.
\end{acknowledgements}

\end{document}